\documentclass[aps,twocolumn]{revtex4}

\usepackage{mathrsfs}
\usepackage{amsmath}
\usepackage{graphicx}
\usepackage{subfigure}
\usepackage{epstopdf}

\newcommand{\be}{\begin{equation}}
\newcommand{\ee}{\end{equation}}

\begin{document}

\title{Magnetization of the QCD vacuum at large fields}

\author{Thomas D. Cohen}
\email{cohen@physics.umd.edu}

\author{Elizabeth S. Werbos}
\email{ewerbos@physics.umd.edu}

\affiliation{Department of Physics, University of Maryland,
College Park, MD 20742-4111}

\begin{abstract}
The response of the QCD vacuum to very large static external magnetic fields ($q B \gg \Lambda_{QCD}^2$) is studied.  In this regime, the magnetization of the QCD vacuum is naturally described via perturbative QCD. Combining pQCD and the Schwinger proper time formalism, we calculate the magnetization of the QCD vacuum due to a strong magnetic field at leading order (one-loop) to be proportional to $B \log B$. We show that the leading  perturbative correction (two-loop) vanishes.
\end{abstract}

\maketitle
\section{Introduction}
Strongly interacting matter is often studied via electromagnetic
probes. For example, the electromagnetic form factors of hadrons give
us information about the distribution and motion of quarks in the
hadron\cite{eff}.  Other useful probes of hadronic---and
nuclear---systems include real and virtual compton
scattering\cite{compton} and various inelastic processes including
those in the deep inelastic regime.  However, it is less commonly
appreciated that--at least in principle---electromagnetism can probe
interesting properties of the QCD vacuum.  In essence an external
electric or magnetic field will polarize the QCD vacuum and the
response of the vacuum to an external quasi-static field gives
significant information about the QCD vacuum.

In practice, the effects of an electromagnetic field on
observables associated with the QCD vacuum are very small for
fields achievable in the laboratory or for known astrophysical sources. For example, the proposed Extreme Light Infrastructure (ELI) project\cite{eli} would produce fields which are several orders of magnitude smaller than needed to have significant effects on the QCD vacuum.  Similarly, the most intense known extended fields in astrophysics---the surface of magnetars---while large enough to cause interesting nuclear physics effects (such as causing the binding of diproton states\cite{dibaryon}) are about two orders of magnitude too small to be of relevance to the QCD vacuum.  Thus the question of how the QCD vacuum responds to extended quasi-static electric or magnetic fields remains largely of theoretical interest.  Nevertheless the question {\it is} of significant interest and has been the subject of study nearly two decades\cite{Klevansky}.

Much of the study of the response of the QCD vacuum to external fields has been in the context of ``QCD-inspired'' models rather than QCD itself.  These include the Nambu and Jona-Lasino (NJL)
model\cite{Gorbar,Klevansky,Gusynin,Klimenko}, the linear $\sigma$
model\cite{Goyal,Sugamura,Schramm}, and a response due to meson loops with the mesons described in a quark model\cite{Kabat}.  All such approaches suffer from the fact that it is difficult to know how well they actually reflect QCD.  Apart from this general concern, there are deep reasons to suspect that models such as the NJL model and the linear sigma models are unlikely to correctly reproduce QCD in interesting regimes.  Calculations in these models are based on mean-field theory which is only known to be justified in the large $N_c$ limit of QCD.  However for weak fields (and small pion masses) the system is near the chiral limit; it is well known that the chiral limit and the large $N_c$ limits do not in general commute\cite{TomLimits} and thus it would not be surprising if the models fail to reproduce the behavior of QCD for weak fields.  As it happens, these models fail to reproduce the qualitative feature of QCD for weak fields\cite{Smilga}.

Fortunately, in the weak-field regime where the mean-field models are particularly  suspect, there is a reliable model-independent approach to the problem---namely, chiral perturbation theory ($\chi$PT)\cite{GasserLeutwyler}.  This approach is based on a scale separation between the masses of the pseudo-Goldstone bosons and the characteristic hadronic scale ($\sim 1$ GeV).  The approach is in essence an expansion in the ratio of light scales (psuedo-Goldstone masses, momenta) to the typical hadronic scale.  The effective theory contains a fixed number of constants at each order in this expansion. In principle, these constants are all determined from QCD.  However, the approach has predictive power even when QCD cannot be fully solved.  By fitting these constants from some finite set of experimental observables, one can predict all other observables---albeit only approximately with an accuracy fixed by the order in the expansion at which one is working.  In the present context, a crucial observation was made in ref.~\cite{Smilga}: an external magnetic field is a source of chiral symmetry breaking and if the magnitude is small ({\it i.e.}, $e H \ll \Lambda_{\rm Hadronic}^2$ where $\Lambda_{\rm Hadronic} \sim 1$GeV is a typical hadronic scale or several times $\Lambda_{QCD}$), its effects can be computed in chiral perturbation theory.   This work considered the limit of massless pions, worked at lowest non-trivial in $\chi$PT and was restricted to purely magnetic fields.  In this regime, the computation is relatively straightforward---it amounts to a one-pion loop calculation for charged pions with the interactions with the external field included to all orders.  Technically this is implemented via the Schwinger proper time formalism\cite{Schwinger}.  Subsequent work extended the analysis to finite mass pions and to the situation of electric as well as magnetic fields\cite{CMW} and to next-to-leading order in $\chi$PT\cite{Werbos}.

The small field regime was tractable precisely because the $q B$ was much smaller than  $\Lambda_{\rm Hadronic}^2$.  For $q B \sim \Lambda_{\rm Hadronic}^2$ chiral perturbation theory breaks down.  In this regime one must either directly solve QCD---for example, via lattice simulations---or resort to models.  The purpose of the present paper is to explore the strong field region: $e H \gg \Lambda_{\rm Hadronic}^2$. In such a regime, there is again a scale separation and one might hope that the this regime is also tractable due to a model-independent and systematic expansion.  Before proceeding it is useful to recall that for the foreseeable future the prospects for obtaining fields in either laboratory or astrophysical settings which are large enough to have noticeable effects on the QCD vacuum remain dim.  The prospects of getting to the strong field regime where the fields are large on hadronic scales is even more remote.

This paper is not the first to focus on this regime: ref. \cite{Kabat} considered strong magnetic fields in the context of a model.  The model has two basic assumptions: i) that the dominant effects can be extracted from two-body interactions between quarks and anti-quarks, and ii) these two-body interactions can be well-approximated by a potential model.  The dynamics of $q \overline{q}$ interactions are altered due to the presence of the magnetic field---which forces the quarks into relativistic Landau orbits and reduces the strong interactions into something which is effectively one dimensional.  At a sufficiently strong field this induces condensation of spin polarized pairs, which yields a magnetization of the vacuum.  In the case of very strong fields it is argued that the system will be tightly bound---in which case the potential can be well approximated by a color coulomb force with a coupling running via perturbative QCD.  This interaction can be solved in the WKB approximation.  A rough estimate for the density of induced pairs is given: it is simply the density at which the (color-singlet) pairs begin to overlap and is thus determined by the size of the pairs.   With these assumptions the magnetization associated with a given flavor of quark is given by
\begin{equation}
\begin{split}
M &\approx \frac{q^2 B \Lambda_{QCD}} {m \pi} \left( \frac{q B}{\Lambda_{QCD}^2} \right)^{\frac{1}{2} \exp (-\pi/2 A))}\\
A & = \frac{8 \pi}{11 N_c - 2 N_f}
\end{split}
\label{Colresult}\end{equation}
where $m$ is the {\it constituent} quark mass and $N_c$ and $N_f$ are the numbers of colors and flavors respectively in the theory.

It is not immediately clear whether the predictions of ref.~\cite{Kabat} are robust.  It was noted in ref.~\cite{Kabat} that quantitatively, the calculation will have non-trivial corrections but it is argued that the qualitative behavior should be reliable.  We merely note here that the dependence of eq.~(\ref{Colresult}) on the constituent quark mass is rather problematic.  In the first place, the notion of a constituent quark mass is a concept which makes sense in the context of models but is not well defined within QCD.  Moreover, the concept becomes particularly problematic in the strong field regime where the typical lengths probed are short---shorter than the characteristic size expected of a constituent quark.

Given this situation, it is important to see whether there is a viable {\it model-independent} way to study the strong magnetic field case.   We argue in this paper that one can apply the techniques of perturbative QCD (pQCD) straightforwardly to this system, provided that one accounts for the strong fields by including the effect of the field on quark propagators to all orders in the field strength---which can be done using a proper time formalism.  The central insight here is that when $q B \gg \Lambda^2$ the characteristic Landau orbit size for a quark is small on the scale of $\Lambda$, and this sets the scale for the coupling when the quark interacts via gluon exchange with other quarks. Due to asymptotic freedom, the quark dynamics in the large field region corresponds to weakly coupled quarks; the principal way the vacuum responds to the field is simply to the rearrangement of states relatively deep in the Dirac sea due to the fields.  In essence, when the characteristic size of a Landau orbit is much smaller than the typical size of a hadron, the quark in responding to the magnetic field doesn't ``know'' about the existence of hadrons and acts to very good approximation as a free particle. Corrections to this picture due to gluon exchange can be handled straightforwardly using perturbative methods.

We focus on the magnetization as our probe of vacuum response.  In the next section we compute the magnetization at leading order (one-loop) perturbative calculation.  Following this we show that at next-to-leading order (two-loops) there is no correction to the one-loop result.  We end with a brief discussion of the implications of these calculations.

\section{Magnetization at leading order}
The magnetization can be calculated as the derivative of the vacuum energy with respect to the magnetic field. In general, the magnetization can be calculated
as\cite{Kabat}
\begin{equation}
{\bf M} = {\bf B}-{\bf H} = \frac{\delta L_{\rm eff}^{matter}}{\delta {\bf B}},
\end{equation}
with $L_{\rm eff}^{matter}$ calculated using the lagrangian appropriate to
our assumptions.

Here we consider the case of $eB \gg \Lambda_{QCD}$ so the natural way to compute the effective Lagrangian
pQCD requires the relevant scales in the problem to be much larger
than $\Lambda_{QCD}$.

The lowest-order effective Lagrangian is obtained from QCD at the one-quark loop level.  The logic here is simply that in pQCD we are doing an expansion in $\alpha_s$ which we take to be small at these scales.  At this order, the various flavors of quark do not communicate with each other---the up quark contribution to the effective exaction at the 
one-loop level is completely independent of down quark properties (charge and mass).   It is only at three-loop order or higher that one has contributions from more than a single flavor.  Thus at this order one can simply compute the contribution from each flavor independently.   We note here that the analysis depends on the mass of the quarks being small.  However, in this context small merely means  that $m_q^2 \ll e B$.  Thus for sufficiently large fields the formalism applies even to heavy quarks.  In intermediate regions, say, $e B \sim 2 GeV^2$, one can compute the perturbative contribution to the magnetization from the u,d and s quarks; in this region contributions from heavier quarks will be suppressed.

The contribution to the effective Lagrangian at this order from a given flavor of quark, $f$,  is that of free fermions in the presence of a constant external magnetic  field; thus it is formally identical to that computed in Schwinger's  classic paper
\cite{Schwinger}, namely:
\begin{equation}
{\cal L}_{\rm eff}^{f} = -\frac{N_c}{8 \pi^2}\int_{0}^{\infty} \frac{ds}{s^3}
e^{-m_{q_f}^2 s}\left(\frac{q_f B s}{\tanh q_f B s} - 1 - \frac{1}{3}(q_f B s)^2\right) \; ,
\end{equation}
where superscript ${\cal L}_{\rm eff}^{f}$ indicates the contribution to the effective action for a particular flavor of quark, $q_f$ is the charge for quarks of that flavor, and $m_{q}$ is the {\it current} quark mass for that flavor.  Changing variables and taking the derivative to get the contribution to the magnetization, we find that at
\begin{equation}
\begin{split}
{\bf M}^{(0)}_f & = -\frac{q_f^2 {\bf B} N_c}{8 \pi^2} I(B,m_{q_f}) \\
I(B,m_{q_f}) & \equiv \int_{0}^{\infty} \frac{dz}{z^2}
e^{-\frac{m_q^2}{q B} z}\left(\coth z - \frac{z}{\sinh^2 z}-\frac{2}{3}z\right),
\end{split}
\end{equation}
where the subscript (0) indicates that it is lowest order in $\alpha_s$.
The $2/3z$ term in the parenthesis has its origin in the renormalization of the electromagnetic field to due fermion loops; its presence ensures that the integrand converges as $z \to 0$.
Because $B \gg m_q^2$, the exponential decays very slowly in the integrand of $I(B,m_{q})$; thus, the large $z$ region dominates.  It is then convenient to separate out the large $z$ contribution; we do this by dividing the integral into high $z$ and low $z$ contributions divided by some number $c$ with the properties that  $c \gg 1$,$\frac{m_q^2}{e B} c \ll 1$:
\begin{equation}
\begin{split}
&I(B,m_{q_f})=I_1 +I_2 \\
I_1 & = \int_{0}^{c}\frac{dz}{z^2} e^{-\frac{m_q^2}{q_f B}z}\left(\coth z - \frac{z}{\sinh^2 z} - \frac{2}{3} z\right)\\
I_2 & = \int_{c}^{\infty}\frac{dz}{z^2} e^{-\frac{m_q^2}{q_f B}z}\left(\coth z - \frac{z}{\sinh^2 z} - \frac{2}{3} z\right).
\end{split}
\end{equation}
In the region of interest, $I_2$ dominates.  Moreover, in this region
 one can drop the $\coth (z)$ and $z/\sinh^2(z)$ terms in $I_2$, since in this region the -2/3$z$ dominates. Thus, up to small corrections,
\begin{equation}
I(B,m_{q}) = - \frac{2}{3}\int_{c}^{\infty}\frac{dz}{z} e^{-m_q^2/(q_f B)z}.
\end{equation}
Evaluating the integral we find that $I(B,m_{q})=-\frac{2}{3}\Gamma^{(0)}\left(\frac{m_q^2}{q_f B} c\right)$ (up to the truncation errors), where $\Gamma^{(0)}(Z)$ is a plica function of zero order.   In the region of interest
($\frac{m_q^2}{q_f B} c \gg 1$), this reduces to:
\begin{equation}
I(B,m_{q_f}) = -\frac{2}{3} \log\left(\frac{m_q^2}{q_f B} c \right )= \frac{2}{3} \log\left(\frac{q_f B}{\mu^2} \right ) + {\rm const}
\label{I} \end{equation}
up to small corrections, where $\mu^2$ is a scale parameter.  The leftover constant in eq.~(\ref{I}) is of the same order as previously neglected terms and can be dropped.  Putting this together yields
\begin{equation}
{\bf M}^{(0)}_f = \frac{q_f^2 {\bf B} N_c}{12 \pi^2}\log \frac{q_f B}{\mu^2} ,
\label{principal} \end{equation}
up to small corrections.  The leading corrections to this are proportional to $B$ itself and, by a judicious choice of $\mu$, these can be fully canceled.  Since important contributions to these canceled terms come from the nonperturbative region, $\mu^2$ encodes important nonperturbative physics.  Note, however, that in the limit of very strong fields, the final answer is very insensitive to the precise value of $\mu$.

Equation (\ref{principal}) is a principal result of this paper.

\section{Magnetization at next-to-leading order}

Having determined the leading order perturbative contribution to the magnetization, it is important to consider corrections.  These may be of two sorts---perturbative and nonperturbative.  The nonperturbative corrections come from low momentum physics and as such cannot be computed directly from QCD via presently known analytic techniques.  However, such effect should be power-law suppressed in $\Lambda_{\rm QCD}^2/(e B)$ compared to the leading-order result of eq. (\ref{principal}) and thus very small at large fields.  One expects the dominant corrections to eq.  (\ref{principal}) to be perturbative.  Such corrections to the magnetization can be written in the form \begin{equation}
{\bf M} = {\bf M}^{(0)} \left ( 1 + c_1 \alpha_s + c_2  \alpha_s^2 + \cdots \right )
\label{pertform} \end{equation} where the $c$'s are dimensionless constants and $\alpha_s$, the strong coupling constant, is evaluated at the scale of the problem: ($e B$).  In this section, we evaluate the leading pertubative correction---$c_1$.   We show that it vanishes.
\begin{figure}[tbp]
\includegraphics{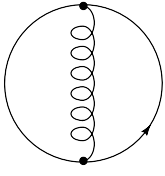}
\caption{\label{fig:vacLoop}
Diagram contributing to magnetization at next-to-leading order}
\end{figure}

The calculation proceeds from the effective action at next-to-leading
order.  This corresponds to the vacuum diagram depicted in
fig.~\ref{fig:vacLoop}.  Note, as with the leading-order effective
actions, the contributions from each flavor are isolated from each
other---each diagram contains only one flavor of quark.  Thus at this
order one can still compute the contribution from each flavor
separately.  The rules for calculating this diagram are identical to
the usual QCD Feynman rules with one important difference: the quark
propagator is replaced by the propagator for a spin-$\frac{1}{2}$
fermion in a constant magnetic field\cite{Schwinger}, and then the
expression for the contribution to the effective Lagrangian at
next-to-leading order for a given flavor can be written as follows:
\begin{widetext}
\begin{equation}
\begin{split}
G^{(B)}_{i j}(p) =& -\delta_{i j}\int_0^{\infty}\frac{ds}{\cos e B s}
\, \exp\left[-i s\left(m_q^2 + p_z^2 + \frac{p_x^2 + p_y^2}{e B s \cot (e B s)} - E^2\right)\right] \\
&\times \Bigg(\left[\cos(e B s) + \gamma_1 \gamma_2 \sin (e B s)\right]\left[\gamma_3 p_z - \gamma_0 E - m_q\right]
+ \frac{\gamma_1 p_x + \gamma_2 p_y}{\cos (e B s)}\Bigg).
\\{\cal L}_{f}^{(1)}(B)  =& -32 \pi^2 \alpha_s N_f^2 N_c \int_0^{\infty}ds\int_0^{\infty}ds' \exp\left[-(s+s')m_q^2\right]I(s,s',B,m_q)\\
I(s,s',B,m_q) =& \frac{1}{(4 \pi)^4 a_1^2 a_2^2 (b_1-b_2)}\left\{-\frac{a_1}{(c c')^2}+a_2 + \log\left(\frac{b_1}{b_2}\right)\left(2 m_q^2 a_1 a_2 + \frac{1}{b_1-b_2}\left[\frac{a_1}{(c c')^2}b_1-a_2 b_2\right]\right)\right\}\\
a_1  =& s + s'\\
a_2  =& s t + s' t'\\
b_1  =& \frac{s s'}{s+s'}\\
b_2  =& \frac{s t s' t'}{s t + s' t'}\\
c =& \cosh (q_f B s), c'=\cosh (q_f B s'), t = \frac{\tanh (q_f B s)}{q_f B s}, t' = \frac{\tanh (q_f B s')}{eq_f B s'}
\end{split}
\end{equation}
\end{widetext}
where the superscript ``1'' indicates an expression at order $\alpha_s^1$.

Note that the expression for the propogator contains the quark mass.  In the high field limit, perturbative expressions should not depend on the quark mass.  We include it here merely to serve as an infrared regulator.

The integral $I(s,s',B,m_q)$ is divergent as written.  However, we are only interested in the $B$ dependence. We can subtract off a divergent constant---the value at $B\to 0$---without affecting the magnetization.  The $B$-dependent part of the total then becomes
\begin{equation}
\begin{split}
{\cal L}_{\rm eff}^{(1)}(B) = &-32 \pi \alpha_s N_f^2 N_c \int_0^{\infty}\!\!\!\!ds\int_0^{\infty}\!\!\!\!ds' \exp\left[-(s+s')m_q^2\right]
\\&\times\left(I(s,s',B,m_q)-I(s,s',0,m_q)\right).
\end{split}
\end{equation}
This integral is convergent; however, it cannot be expressed analytically in closed form. Fortunately, we are interested in the regime of $e B \gg
m_q^2$.  Thus, we can implement the same trick as was employed for the one-loop expression: divide the integral into a high and low $s$ region with the knowledge that the dominant contribution comes from a large $s$ region: we obtain
\begin{equation}
\begin{split}
{\cal L}_{f}^{(1)}(B) = &-32 \pi \alpha_s N_f^2 N_c (q_f B) \frac{m_0^2}{2(4\pi)^4}\textrm{Ei}(-2 \frac{m_q^2}{m_0^2})
\\&\times\left(2 -2 e^{-\frac{m_q^2}{m_0^2}}+\left(\frac{m_q^2}{m_0^2}\right)\Gamma(0,\frac{m_q^2}{m_0^2})\right),
\end{split}
\end{equation}
up to small corrections, where $m_0$ is the separation scale introduced. Working in the regime $m_0 \gg m_q$ and taking the derivative with $B$, we find a dependence of the form
\begin{equation}
{\bf M}_f^{(1)}(B) =  \pi \alpha_s  N_c (q_f^2 {\bf B})\frac{m_q^2}{q_f B}\frac{1}{16 \pi^4}\left(\log \frac{e B}{\mu^2}\right)^2,
\end{equation}
where $\mu$ is some renormalization scale chosen to minimize corrections due to nonperturbative effects. Combining this with the lowest-order result, yields
\begin{equation}
\begin{split}
{\bf M}_f(B) = &q_f^2{\bf B}\frac{N_c}{12\pi^2}\log \frac{q_f B}{\mu^2}
\\&\times\left(1 +  4 \pi \alpha_s  \frac{m_q^2}{q_f B}\frac{3}{16\pi^2}\log \left(\frac{q_f B}{\mu^2}\right)\right),
\label{ff}
\end{split}
\end{equation}
plus corrections of order $\alpha_s^2$.

The critical thing to notice is that the correction term in eq.~({\ref{ff}) is proportional to $\alpha_s m_q^2/(q_f B)$ times the lowest-order term rather than just a constant times $\alpha_s$.  Thus, it is power-law suppressed in $B$ and the coefficient $c_1$ is zero.  In fact, for light quarks this correction term is simply not reliable---it  represents a small piece of the nonperturbative power-suppressed contribution and cannot be separated from it in a meaningful way.  For the case of heavy quarks with very strong fields (the regime $q_f B \gg m_q^2 \gg \Lambda_{QCD}^2$) the correction term is presumably reliable: it still represents a nonperturbative correction that is power-law suppressed at strong fields, but it is the leading contribution to this in a heavy quark expansion.  Regardless of whether the quark masses are small or large, eq.~(\ref{principal}) is accurate up to corrections of order $\alpha_s^2$.

\section{Conclusions \label{conc}}

We have examined the magnetization of the QCD vacuum in the presence
of a constant magnetic field in the strong field regime, $e B \gg
\Lambda_{QCD}$, where perturbative QCD should be valid.  We have found the contribution from any flavor of quark with $q_f B \gg m_q^2$.  Combining these together, one finds that, in a regime in which the $e B$ is either much greater or much smaller than all of the quark masses, the magnetization can be written as
\begin{equation}
{\bf M}(B) = \left (\sum_{\substack{{\rm active}\\{\rm flavors}}} \frac{q_f^2}{e^2} \right ) {\bf e^2 B}\frac{1}{4 \pi^2}\log \frac{e B}{\mu^2}\left(1 +  {\cal O} (\alpha_s^2)   \right),
\label{final}\end{equation}
where active flavors refer to those whose squared mass is well below $e B$.

One striking feature of this result is that the two-loop correction makes no perturbative contribution.  At present, we are unsure as to whether this indicates something deep about the underlying structure of the theory, or rather it is something of an accident.

Much of the motivation for this work was to obtain a model-independent prediction for the strong field region.  Thus it is important to explore the relationship of our result to eq.~(\ref{Colresult}) obtained in ref.~\cite{Kabat} using a plausible model and heuristic reasoning.  At first glance, the two results are rather similar once one recognizes that the constituent quark mass is of $O(\Lambda)$ in eq.~(\ref{Colresult}): both calculations have the magnetization growing linearly with $B$ times a slowly increasing function of $B/\Lambda$---in the case of eq.~(\ref{final}), logarithmically (noting that $\mu \sim \Lambda$), and in the case of eq.~(\ref{Colresult}), a {\it very} small power law.  Thus, it might appear that these are two essentially complementary descriptions of essentially the same physics.  However, when one looks closely, there are important qualitative differences between the two descriptions.

From a theoretical perspective, one can use the number of colors as a probe of the nature of the dynamics\cite{Witten,tHooft}.  Note that the perturbative expression derived in eq.~(\ref{ff}) has a magnetization proportional to $N_c$.  The origin of this is clear.  The active players in dynamics are quarks---which act nearly as free particles for the purposes of computing the magnetization, since the length scale of Landau orbits is small compared to characteristic hadronic sizes.  Since the number of quarks participating is proportional to $N_c$, so is the magnetization.  In contrast,  eq.~(\ref{Colresult}) has no overall factor of $N_c$.  This reflects the dynamics modeled in ref.~\cite{Kabat}, where the magnetization was driven by a condensation of  color-singlet ``mesons''. Moreover, phenomenologically, the two results differ qualitatively in the regime of {\it extremely} large fields.  In such a regime the power-law growth in  eq.~(\ref{Colresult}) ultimately becomes much larger than the logarithm in eq.~(\ref{final}).

We believe that in the extreme strong field regime, the perturbative calculation described here should become increasingly accurate for reasons discussed above: at such extreme fields the localization scale of a Landau orbit is far smaller than the size of hadrons and, as far as vacuum polarization effects are concerned, should appear to be essentially free.  Thus, we suspect that the model described in ref.~\cite{Kabat} must break down by the time the fields get very strong.  This is highly plausible.  The model depends on the notion of constituent quarks.  This concept, while of possible utility in describing low-lying hadrons, is not appropriate for describing the perturbative regime.

Support of the U. S. Department of Energy under grant number DE-FG02-93ER-40762 is gratefully acknowledged.

\end{document}